\documentclass[11pt]{article}
\usepackage{amsfonts}
\usepackage{mathrsfs}
\usepackage[latin1]{inputenc}
\usepackage{float}
\usepackage{amsmath,amssymb}
\usepackage{amsfonts}
\usepackage{latexsym}
\usepackage{epstopdf}
\usepackage{geometry} 
\usepackage{bbm} 
\usepackage[active]{srcltx}
\usepackage{graphicx}
\usepackage{epstopdf}
\usepackage{booktabs}
\usepackage{multirow}
\usepackage{siunitx}

\usepackage{setspace}
\usepackage[
bookmarks=true,         
bookmarksnumbered=true, 
colorlinks=true, pdfstartview=FitV, linkcolor=blue, citecolor=blue,
urlcolor=blue]{hyperref}

\providecommand{\U}[1]{\protect\rule{.1in}{.1in}}
\newtheorem {theorem}{Theorem}[section]

\newtheorem{lemma}{Lemma}[section]

\newtheorem{remark}{Remark}[section]

\newcommand{\E}{\mathbb{E}}

\newcommand{\abs}[1]{\lvert#1\rvert}

\newcommand{\bi}[1]{\mbox{\boldmath{$ #1 $}}}

\begin{document}

\title{Grouped feature screening for ultrahigh-dimensional classification via Gini distance correlation}
\vspace{0.5cm}
\author{Yongli Sang\textsuperscript{a} and Xin Dang\textsuperscript{b}\thanks{CONTACT: Xin Dang; Email: xdang@olemiss.edu}}
\date{%
  \textsuperscript{a}Department of Mathematics, University of Louisiana at Lafayette, Lafayette, LA 70504, USA\\
  \textsuperscript{b}Department of Mathematics, University of Mississippi, University, MS 38677, USA\\[2ex]%
     \today
}

\maketitle

\begin{abstract}
Gini distance correlation (GDC) was recently proposed to measure the dependence between a categorical variable, $Y$, and a numerical random vector, $\bi X$. It mutually characterizes independence between $\bi X$ and $Y$. In this article, we utilize the GDC to establish a feature screening for ultrahigh-dimensional discriminant analysis where the response variable is categorical.  It can be used for screening individual features as well as grouped features. The proposed procedure possesses several appealing properties. It is model-free. No model specification is needed. It holds the sure independence screening property and the ranking consistency property.  The proposed screening method can also deal with the case that the response has divergent number of categories. We conduct several Monte Carlo simulation studies to examine the finite sample performance of the proposed screening procedure. Real data analysis for two real life datasets are illustrated.

 \end{abstract}
\noindent {\bf Keywords:} Discrimination analysis; Gini distance correlation; Group screening;  Feature screening; Ranking consistency property; Sure independence screening; Ultrahigh dimensionality.
\noindent 

\vskip.2cm 
\noindent  {\textit{MSC 2020 subject classification}: 62H30 }

\section{Introduction}
In the last decade, independence feature screening methods have been considered to be efficient for ultrahigh-dimensional data.  The sure independent screening (SIS) \cite{Fan08} ranks the importance of individual variable by its Pearson correlation with the response.  In the linear model setting,  the procedure has been shown to have the sure screening property, meaning that the probability of all the important variables to be selected tending to one.  Since this seminal paper, various screening methods have been proposed \cite{Fan10, Fan11, Fan09, Zhu11}. Hall and Miller \cite{Hall09} utilized a generalized correlation to do screening.  The generalized correlation is defined as the maximum correlation of transformations for the feature variable with the response.  If the correlation is maximized among transformations of feature variable and also response transformations, the maximal correlation is obtained and is used for screening in \cite{Huang16}.  Kendall's $\tau$ correlation \cite{Kendall38,Kendall90} is used in \cite{LP12} for the robustness propose.  Distance correlation is applied in \cite{Li12} for the model-free property and closely related one is the martingale difference correlation used in \cite{Shao14}.  All of above mentioned procedures consider a numerical response. 

When the response is categorical, several screening methods have been proposed \cite{Cui15, Ni16, Cheng17,Lai17}.  Cui, Li and Zhong \cite{Cui15} used difference between the marginal and the conditional distribution function to construct an index for feature screening.  
In binary classification,  the index is reduced to the Cram\'{e}r-von Mise test statistic and its rank-based representation is studied in \cite{Curry18}.   A rank based screening procedure based on conditional expectation of the rank of predictor was proposed by Cheng et al. \cite{Cheng17}. He, Ma and Xu  \cite{He19} modified the procedure of \cite{Cui15} and considered the Anderson-Darling type index that provides more weights on the observations on the tails. 
 In \cite{Ni16}, entropy is used to measure dependence of each predictor with the categorical response.  All these feature screening procedures are based on the individual predictors and they are not able to incorporate structured predictors like grouped variables where all the covariates are highly correlated. 
In many cases of ultrahigh-dimensional data, predictors are naturally grouped. 

 The results for grouped feature screening for ultrahigh-dimensional data are limited. Niu $et$ $al.$ \cite{Niu2020} proposed a marginal group sure screening method in linear models by ranking the magnitude of marginal estimators based on each grouped predictor. Qiu and Ahn \cite{Qiu2020} developed three new screening methods, gSIS (group sure independence screening), gHOLP (group high-dimensional ordinary least-squares projector), and gAR2 (groupwise adjusted R-squares screening).  Song and Xie \cite{Song2019} proposed a group screening procedure via a $F$-test statistic by extending the original sure independence screening procedure when the group information is known.  All the above mentioned methods mainly focus on linear models.  For classification, three model-free grouped screening methods were proposed recently. In \cite{He2022} and \cite{Wang22},  entropy and mutual information are used to measure the prediction importance of each grouped covariate.  In \cite{Wang23}, Gini impurity as a classification index is used for screening measure.  However, those procedures are designed for categorical covariates. When covariates are numerical, they have to reply on slicing in order to change the numerical covariates into categories, and hence lose a quite amount of information. Also how to slice is a big issue to apply those methods. 
 
 In this paper, we propose a new  model-free grouped screening method for discriminant analysis  in ultrahigh-dimensional data. Although the proposed method can deal with categorical covariates, it especially is useful for screening numerical covariates, which is the most common case for classification. The new procedure is based on the Gini distance correlation (GDC) between the group of covariates with the response. The GDC is a recently proposed dependence measure of a categorical variable, $Y$, and a $p$-variate random vector $\bi X$ \cite{Dang2021}. It has been shown that the GDC between $\bi X$ and $Y$ equals to zero if and only if $\bi X$ and $Y$ are independent.  Such independence indicates that the predictors $\bi X$ have the same statistical distributions under different levels of the categorical variable $Y$. This  enables us to construct a new index for feature screening in ultrahigh-dimensional data.  The proposed screening procedure ranks the importance of group predictors by the GDC of the group covariates with the response. In GDC, the dimensionality of the covariates can be arbitrary so the proposed procedure can be applied to both the individual feature screening and the grouped feature screening. 
 
The proposed screening procedure not only holds the sure screening property and the ranking consistency property but also possesses the following desirable merits.  
 \begin{enumerate}
 \item 
It is model-free. That is, it does not require model specification for the response or predictors. 
\item The new procedure can be applied to screen both individual predictor and grouped predictors. 
\item It also allows a divergent number of levels for the response variable. 
\end{enumerate}

The remainder of the paper is organized as follows. In Section \ref{sec:fc}, we  first introduce the GDC and then establish the sure independence screening procedure. The presentation is majorly for the group screening and treats the individual screening as the special case.  
 In Section \ref{sec:simulationstudy}, we conduct simulation studies to evaluate the performance  of the proposed screening procedure.  Real data analysis for two datasets is illustrated in Section \ref{sec:realdata} to compare the proposed procedure with current existing approaches. 
We conclude and discuss some connection with an existing method in Section \ref{sec:conclusion}. All technical proofs are provided in Appendix.

\section{Grouped feature screening}\label{sec:fc}

Let $Y$ be the categorical response variable taking values $L_1, ...,L_K$ with $K\geq 2$ and $\bi X=(X_1, ..., X_p)^T$ be the predictor vector. Suppose that based on some domain knowledge, the covariates $\bi X$  are partitioned into $r$ groups of predictors $(\bi X_{(1)},  \bi X_{(2)},..., \bi X_{(r)})$, where the $l^{th}$ group is $\bi X_{(l)} = (X_{l1},...X_{lq_l})^T$ with dimensionality  $q_l$ satisfying  $q_1+q_2+...+q_r=p$. We assume that $q_l$, $l=1,...,r,$ are finite.   In an ultrahigh-dimensional setting, the group number $r$ (and hence $p$) greatly exceeds the sample size $n$. It is thus natural to assume that only a small number of predictor groups are relevant to $Y$.  The goal of a screening procedure is to select a reduced model with a moderate scale that almost fully includes the relevant predictor groups. 
Before present details of the screening procedure, we review the Gini distance correlation between $\bi X_{(l)}$ and $Y$.

\subsection{Gini distance correlation}\label{subsec:gdc}
Let the distribution of $Y$ be $P(Y=L_k)=p_k >0$ for $k=1,2,...,K$. Assume the distribution of $\bi X_{(l)}$ is $F_l$ and the conditional distribution of $\bi X_{(l)}$ given $Y=L_k$ is $F_{l|k}$. Denote $\psi_{l|k}$ and $\psi_l$ as the characteristic functions of $F_{l|k}$ and $F_{l}$, respectively.
Then the Gini distance covariance between $\bi X_{(l)}$ and $Y$ is defined as  
\begin{equation}\label{Gcov}
\mbox{gCov}(\bi X_{(l)},Y)  = c(q_l)\sum_{k=1}^K p_k  \int_{\mathbb{R}^{q_l}} \frac{|\psi_{l|k}(\bi t) -\psi_l(\bi t)|^2}{\|\bi t\|^{q_l+1}} d\bi t,
\end{equation}
where $c(\cdot)$ is a known constant depending on dimension $q_l$, 
$\|\bi t\|$ is the Euclidean norm of $\bi t \in \mathbb{R}^{q_l}$ and $|\psi|^2 = \psi \bar{\psi}$ for a complex function $\psi$.  
The categorical Gini covariance measures dependence based on the weighted (squared) distance between marginal and conditional characteristic functions.  
We see that gCov$(\bi X_{(l)},Y) \geq 0$ and gCov$(\bi X_{(l)},Y) = 0$ mutually implies independence of $\bi X_{(l)}$ and $Y$. 

When $q_l =1$, the Gini distance covariance between $Y$ and univariate $X_{(l)}$ defined in (\ref{Gcov}) can be simplified as
\begin{equation*}\label{Gcov_1}
\mbox{gCov}( X_{(l)},Y)  = \sum_{k=1}^K p_k  \int_{\mathbb{R}} \big(F_{l|k}(x)-F_l(x)\big)^2 dx.
\end{equation*}
It is the weighted squared distance between the marginal distribution and the conditional distribution. It is closely related to the mean variance (MV) index used in \cite{Cui15}. That is,
\begin{equation*}\label{MV}
\textrm{MV}( X_{(l)},Y)  = \sum_{k=1}^K p_k  \int_{\mathbb{R}} \big(F_{l|k}(x)-F_l(x)\big)^2 d F_l(x), 
\end{equation*}
which is a weighted Cram\'{e}r-von Mises distance between the marginal and conditional distributions. The MV only applies for one dimension and it has no direct multivariate generalization, while the Gini covariance works for univariate and multivariate cases.  

The Gini covariance between $\bi X_{(l)}$ and $Y$ in (\ref{Gcov}) can be represented by multivariate Gini mean differences (GMD) \cite{Dang2021}. That is,
\begin{equation}\label{Gcov_2}
\mbox{gCov}(\bi X_{(l)},Y) = \Delta_l-\sum_{k=1}^Kp_k\Delta_{l|k},
\end{equation}
where 
$\Delta_l =\E\|\bi X_{(l)1}-\bi X_{(l)2}\|$  is the GMD for $F_l$ and
$\Delta_{l|k}=\E \|\bi X_{(l|k)1}-\bi X_{(l|k)2}\|$ denotes the GMD of $F_{l|k}, k=1,..., K$ with $(\bi X_{(l1)},\bi X_{(l2)})$ and $(\bi X_{(l|k)1}, \bi X_{(l|k)2})$ being independent pair variables independently from $F_l$ and $F_{l|k}$, respectively.   This representation (\ref{Gcov_2}) provides a more direct way than (\ref{Gcov}) to estimate the Gini covariance. Also the corresponding Gini distance correlation (GDC) can simply be defined by 
\begin{equation} \label{mgc}
\rho_l= \frac{ \Delta_l-\sum_{k=1}^Kp_k\Delta_{l|k}}{\Delta_l}.
\end{equation}
The Gini distance correlation (\ref{mgc}) has a nice interpretation as the ratio of the between variation and the total variation. 
It has a range of $[0,1]$.  It is zero if and only if numerical variable $\bi X_{(l)}$ and categorical variable $Y$ is independent. Those remarkable properties motivate us to use the GDC for ranking the importance of $\bi X_{(l)}$ in classification.     

Suppose that  ${\cal D} = \big\{\big(\bi X_1, Y_1\big), \big(\bi X_{2}, Y_2\big), ...., \big(\bi X_{n}, Y_n\big)\big\}$ be a random sample from the joint distribution of $\bi X$ and $Y$. Let ${\cal D}_{l}= \big\{\big(\bi X_{(l)i}, Y_i\big)\} _{i=1}^n$ be the sample data only containing the $l^{th}$ group of predictors.  We write ${\cal D}_{l} ={\cal D}_{l|1}\cup {\cal D}_{l|2}...\cup {\cal D}_{l|K}$,  where ${\cal D}_{l|k}=\left \{\bi X_{(l|k)1}, \bi X_{(l|k)2}, ...,\bi X_{(l|k)n_k}\right \}$ is the subsample of ${\cal D}_{l}$ with $Y=L_k$ and $n_k$ is the number of sample points in the $k^{th}$ class. Then the Gini distance correlation in (\ref{mgc}) can be estimated by 
\begin{align}\label{gcovn}
\hat{\rho}_l= \dfrac{{n \choose 2}^{-1}\sum_{1 \leq i <j \leq n}\|\bi X_{(l)i}-\bi X_{(l)j}\| -  \sum_{k=1}^K \hat{p}_k {n_k \choose 2}^{-1}\sum_{1 \leq i<j \leq n_k}\|\bi X_{(l|k)i}-\bi X_{(l|k)j}\|}{{n \choose 2}^{-1}\sum_{1 \leq i <j \leq n}\|\bi X_{(l)i}-\bi X_{(l)j}\|},
\end{align}
where $\hat{p}_k=\dfrac{n_k}{n}= \dfrac{1}{n}\sum_{i=1}^n \mathbbm{1}(Y=L_k), k=1,...,K$ with  $\mathbbm{1}(\cdot)$ the indicator function.

The computation of (\ref{gcovn}) is straightforward with the computational complexity $O(n^2)$ in general.  However, for univariate $X_{(l)}$ with dimension $q_l =1$, a simple fast algorithm for (\ref{gcovn}) only costs $O(n\log n)$. This makes the marginal feature screening appealing even for large $n$. Dang $et$ $al.$ \cite{Dang2021} and Sang and Dang \cite{Sang2022} have  explored the properties of the GDC in low-dimensional and high-dimensional settings, respectively.  Compared with the popular distance correlation \cite{Szekely07, Li12}, the GDC is more straightforward to perform statistical inference and it is more robust to deal with unbalanced data. We expect that those nice properties of the GDC will be inherited by the feature screening method for ultrahigh-dimensional data.     

\subsection{GDC based grouped feature screening}
In this section, we propose an independence screening procedure built upon the GDC. 
Denote by $G(Y=L_k|\bi X)$ the conditional distribution of $Y$ given $\bi X$. With no need to specify a classification model, we define the index set of the active and inactive groups respectively  by 
\begin{align*}
&{\cal A} = \{ l: G(Y=L_k|\bi X) \mbox{ functionally depends on }  X_l \mbox{ for some } k \}, \nonumber\\
&{\cal I} = \{ l: G(Y=L_k|\bi X) \mbox{ does not functionally depends on } X_l \mbox{ for all }k\}. 
\end{align*}

Define $\bi X_{\cal{A}}=\{\bi X_{(l)}: l \in \cal{A}\}$ and $\bi X_{\cal{I}}=\{\bi X_{(l)}: l \in \cal{I}\}$, and refer to $\bi X_{\cal{A}}$ as an active predictor vector and its complementary 
$\bi X_{\cal{I}}$ as an inactive predictor vector. The index subset $\cal{A}$ of all active predictors or, equivalently, the index subset $\cal{I}$ of all inactive predictors, is the objective of our primary interest. The definition of $\cal{A}$ and $\cal{I}$ implies that $\bi X_{\cal{I}}$ are independent of $Y$ given $\bi X_{\cal{A}}$.
To identify the set ${\cal A}$, we use the Gini correlation between $\bi X_{(l)}$ and $Y$.  

We rank  the importance of grouped predictors $\bi X_{(l)}$ by ranking the GDC of $\bi X_{(l)}$ with $Y$, $\rho_l, l=1,...,r$.  Note that $\rho_l = 0$ if and only if $\bi X_{(l)}$ and $Y$ are independent. $\rho_l $ is an effective index to separate the active and inactive groups. For example, under the partial orthogonality condition \cite{Fan10} in which $\{ \bi X_{(l)}: l \in {\cal{A}} \}$ are independent of $\{ \bi X_{(l)}: l \in {\cal{I}} \}$, then $\rho_l > 0$ if $l \in {\cal{A}}$ and $\rho_l=0$ for $l \in \cal{I}$. Also, 
the GDC does not need model assumption and allows for arbitrary dimensionality of predictors. So it can be applied to screen individual covariate or grouped covariates. 

We select a set of important groups with large $\hat{\rho}_j$. That is, we estimate ${\cal A}$ by $\hat{\cal A}$, which is 
$$
\hat{\cal A} = \{ l: \hat{\rho}_l \geq c n^{-\kappa}, \mbox{ for } 1 \leq l \leq r\},
$$
where $c>0$ and $\kappa \in [0,1/2)$ are predetermined thresholding values defined in condition \textbf{C3}. In practice, for a given size $d<n$, we can select a reduced  model:
$$
\hat{\cal A}^{*} = \{ l: \hat{\rho}_l \  \text{is among the top $d$ largest of all}\}.
$$

With some conditions on the minimum Gini correlation of active group of predictors, e.g.,  $\min_{l \in {\cal A}} \rho_l  \geq 2 c n^{-\kappa}$, we would like to prove the sure screening property, which can be established by deriving universal exponential convergence bounds.

The following conditions are imposed to facilitate the technical proofs, although they may not be the weakest ones. 

\begin{description}
\item[C1] There exist two positive constants $c_1$ and $c_2$ such that 
\[
c_1/K_n \leq \min_{1 \leq k \leq K_n} p_k \leq \max_{1 \leq k \leq K_n} p_k \leq c_2/K_n.
\]  Assume that $K_n=O(n^{\tau})$ for $\tau \geq 0$.
\item[C2] $\bi X$ satisfies the sub-exponential tail probability uniformly in dimension $p$ and in class. That is, there exists a positive constant $s_0$ such that for all $0 \le s\leq 2s_0$, 
\[
\sup_r \max_{1\leq l \leq r} \sup_K \max_{1\leq k \leq K} \E \{\exp(s\|\bi X_{(l|k)}\|)\} < \infty.
\] 

\item[C3] The minimum Gini correlation of an active predictor group satisfies 
\[
 \min_{l \in \cal{A}} \rho_l  \geq 2cn^{-\kappa},\,\,\, \mbox{ for some constants } c>0 \mbox{ and } 0\leq \kappa <1/2. 
\]
\end{description}

Condition \textbf{C1} requires that the proportion of each class of the categorical variable $Y$ cannot be either too small or too large, meaning that no class dominates over others.  Note that the number of classes $K_n=O(n^\tau)$ allows to be diverging to the infinity.  The sub-exponential condition \textbf{C2} can be satisfied directly if $\bi X$ is uniformly bounded or follows a multivariate normal distribution. It is weaker than the sub-Gaussian condition used in \cite{Li12} and it permits more heavy tailed distributions~\cite{Vershynin18}.  Assumption \textbf{C3} is a typical condition in sure independence screening procedure and it requires the GDC of an active of grouped predictors cannot be too small. It is of order $n^{-\kappa}$ that admits the minimum
true signal to vanish to zero as the sample size $n$ approaches the infinity.

We establish the sure screening property for the grouped screening procedure in the following theorem. 
\begin{theorem}\label{ucb_gCov}
Under conditions \textbf{C1-C2}, for any $0<\beta<1/2-\kappa$ and $0 \leq \tau< 1-2 \kappa$, there exist positive constants $b_1$,  $b_2$, $b_3$ depending on $c$, $c_1$, and $c_2$, such that  
\begin{align}
&P(\max_{1 \leq l \leq r}|\hat{\rho}_l-\rho_l| \geq c n^{-\kappa})\leq \nonumber \\
&   O\left\{r\left[  \exp(-b_1 n^{1-2(\kappa+\beta)}+\tau \log n)+  \exp(-b_2 n^{\beta}+(1+\tau)\log n)+ \exp(-b_3 n^{1-2 \kappa-\tau}+\tau\log n)\right ]\right\}. \label{uni_con}
\end{align}
Under conditions \textbf{C1-C3}, we have
\begin{align}
 &P({\cal A} \subseteq \hat{\cal A})\geq  \nonumber \\
& 1-O\bigg(d_n\left[  \exp(-b_1 n^{1-2(\kappa+\beta)}+\tau \log n)+  \exp(-b_2 n^{\beta}+(1+\tau)\log n)+ \exp(-b_3 n^{1-2 \kappa-\tau}+\tau\log n)\right ]\bigg), \label{screening}
 \end{align}
where $d_n = | {\cal A}|$, the cardinality of ${\cal A}$.
\end{theorem}
This demonstrates that our GDC based screening procedure can be applied for Non-Polynomial (NP) dimensionality. We mean that $\log p \geq \log r=O(n^{\omega})$, where $\omega< \min\{1-2(\kappa+\beta), \beta, 1-2\kappa-\tau\}$ with $0 \leq \kappa <1/2$, $0 \leq\beta<1/2-\kappa$ and $0 \leq \tau <1-2\kappa$, which depends on the minimum true signal strengthen and the number of response
classes.

If $K_n$ is fixed with $\tau=0$, the results for Theorem \ref{ucb_gCov} can be improved as 
\begin{align} \label{fixK}
P(\max_{1 \leq l \leq r}|\hat{\rho}_l-\rho_l|\geq c n^{-\kappa}) & \leq  O\left\{r\left[  \exp(-b_1 n^{1-2(\kappa+\beta)})+  \exp(-b_2 n^{\beta}+\log n)\right ]\right\}.
\end{align}
So we can handle even larger NP-dimensionality where $\log p \geq \log r=O( n^\omega)$, where  $\omega< \min\{1-2(\kappa+\beta), \beta\}$ with $0 \leq \kappa <1/2$ and $0 \leq\beta<1/2-\kappa$. With a balance of two terms of (\ref{fixK}), we  choose the optimal order $\beta = (1-2\kappa)/3$, then for some constant $b >0$, 
\begin{align*}
P(\max_{1 \leq l \leq r}|\hat{\rho}_l-\rho_l|\geq c n^{-\kappa}) & \leq  O\left\{r\left[   \exp(-b n^{(1-2\kappa)/3}+\log n)\right ]\right\}.
\end{align*}

Another property for independence screening is the ranking consistency property considered by Zhu $et$ $al.$ \cite{Zhu11}. To investigate the ranking consistency property, we additionally assume the following condition:
\begin{description}
\item[C4] For some positive constant $c_3$,
\[
\displaystyle{\lim\inf_{r \to \infty}}\{\min_{l \in \mathcal{A}}\rho_l-\max_{l \in \mathcal{I}}\rho_l\}\geq c_3.
\]
\end{description}
Condition C4 imposes the separability of active and inactive groups in terms of the population Gini distance correlation.   It is easy to see that C4 is weaker than the partial orthogonal condition. This is because if $\{ \bi X_{(l)}: l \in {\cal{A}} \}$ are independent of $\{ \bi X_{(l)}: l \in {\cal{I}} \}$, then $\rho_l > 0$ if $l \in {\cal{A}}$ and $\rho_l=0$ for $l \in \cal{I}$, implying that C4 holds. 

\begin{theorem}\label{rc}[Ranking Consistency Property]
If conditions \textbf{C1}-\textbf{C4} hold for $\dfrac{K_n \log n}{n^\alpha}=o(1)$ and $\dfrac{K_n \log p}{n^{\alpha}}=o(1)$, for $\alpha=\min\{1-2\beta+\tau, \beta+\tau, 1\}$ then we have that
\[
{\lim \inf}_{n \to \infty}\{\min_{l \in \mathcal{A}}\hat{\rho}_l-\max_{l \in \mathcal{I}}\hat{\rho}_l\}\geq 0, a.s.
\]
\end{theorem}

The result in Theorem \ref{rc} is stronger than the sure screening property. It shows that with high probability the sample GDC of active predictors are always ranked beyond those of inactive ones.  With an ideal thresholding value, the proposed screening procedure is able to identify the active predictor groups based on sample data as long as active and inactive groups are separable in the population level.

\begin{remark}
When $r=p$ with $q_l =1$ for all $l=1,..., r$, the screening procedure is for marginal screening. We rank each individual covariate by its marginal GDC with the response variable. The results of Theorem \ref{ucb_gCov} and Theorem \ref{rc} hold for the marginal screening.
\end{remark}

\begin{remark}
Zhang et al. (2019)~\cite{Zhang19} extended Gini distance covariance and
correlation to reproducing kernel Hilbert spaces (RKHS). With a choice of a bounded kernel, the screening procedure based on the kernel Gini correlation is able to relax the condition C2, but still possesses the sure screening property and the rank consistency property. The uniform exponential convergence rate of the sample kernel Gini correlation in their Corollary 7 is ready to establish the sure independence screening property. 
\end{remark}

\section{Simulation}\label{sec:simulationstudy}
We consider four simulation studies to evaluate performance the proposed screening method, denoted as GN. For the individual screening, GN is compared with the following four methods.

\begin{itemize}
\item $R^2$: Pearson $R^2$ in univariate ANOVA;
\item MV: Mean-Variance  method proposed in~\cite{Cui15};
\item AD:  Anderson Darling type Mean-Variance method considered in~\cite{He19};
\item DT: screening procedure based on the distance correlation~\cite{Li12, Dang2021}. 
\end{itemize} 
Note that all methods can be viewed as ranking predictors by dependence measures between categorical and numerical variables. The distance method (DN) is based on the set difference metric for the categorical variable.   More details refer to \cite{Dang2021}.  

As GN, DN is also used for group screening. They are compared with the other three group screening methods. 
\begin{itemize}
\item EN:  entropy information gain method proposed in~\cite{He2022};
\item ENR:  entropy information ratio method used in~\cite{Wang22};
\item GI:  Gini impurity method introduced in~\cite{Wang23}.
\end{itemize} 

The minimum model size (MMS) of models including all active predictors or all active predictor groups are applied to measure the effectiveness of each screening approach. 
For each example, we repeat the procedure $M=500$ times,  and compute the median and the robust estimation of the standard deviation (IQR/1.34, denoted by RSD) of MMS.  In addition, denoted as $P_l$, the proportion including a single active predictor $X_l$ or an active predictor group $\bi X_{(l)}$  is computed and the proportion including all active predictors, denoted by $P_{\textrm{all}}$ is reported for a given model size $d = [n/ \log n]$, where $n$ is the sample size and $[x]$ denotes the integer part of $x$. $P_{\textrm{min}}$ and $P_{\textrm{max}}$ are the minimum and maximum of $P_l$'s, respectively. \\

\noindent{\bf \textit{Example 3.1 (Ultrahigh Dimensional Linear Discriminant Analysis)}}

We consider a linear discriminant
analysis (LDA) problem with ultrahigh dimensional predictors by following
the similar settings in Cui,  Li and Zhong \cite{Cui15} and Pan, Wang, and Li \cite{Pan13}. We consider $K=3$ and $K=10$ classes. For each  $i^{th}$ observation, the categorical response $Y_i$ is generated
from three different discrete distributions of $K$ categories with $\bi p = (p_1,p_2,...p_K)$ where $p_k = P(Y_i=L_k)$: 
(I) balanced,  $p_k = 1/K$ with $k = 1, . . . , K$;  (II) slightly unbalanced,  $\bi p = (3/12, 4/12, 5/12)$ for $K=3$ and  $\bi p  =(6,7,8,9,10,10,11,12,13,14)/100 $ for $K=10$;  
(III) heavily unbalanced, $\bi p = (0.1,0.3,0.6)$ for $K=3$ and  $\bi p =(2,4,6,8,10,10,12,14,16,18)/100 $ for $K=10$. 
This is designed for examining how class imbalance and class number effect performance of each marginal screening screening method. 

Given $Y_i = L_k$, the $i^{th}$ predictor $\bi X_i$ is then generated by letting $\bi X_i = \bi \mu_k + \bi \epsilon_i$ , where the mean vector  $\bi \mu_k = (\mu_{k1}, \mu_{k2},....\mu_{kp})^T$ is a $p$-dimensional
vector with the $k^{th}$ element $\mu_{kk} = 3$ but others are all zero, and $\bi \epsilon_i = (\epsilon_{i1}, \epsilon_{i2},...\epsilon_{ip})^T$  is a $p$-dimensional error term.
Here, we consider two cases of the error term: (1) $\epsilon_{ij}$'s are iid from $N(0, 1)$; 
(2) $\epsilon_{ij}$'s are iid from $t(2)$.  The Case (2) is for
examining the robustness of an independence screening
method. We consider $p =2000$ predictors in which $ X_1, X_2,  \text{and}  \ X_3$ are the active predictors for $K=3$ and $X_1,  X_2, ..., \text{and}  \ X_{10}$ are the active predictors when $K=10$.    The sample sizes are $n=60$ for $K=3$, and $n=200$ for $K=10$, respectively.  
The median and RSD of MMS as well as the proportion of containing each active predictor  and all active predictors are reported in Table \ref{tab:example1}. In order to save space, we just report $P_{\textrm{min}}$ and $P_{\textrm{max}}$ instead of $10$ proportions for $K=10$.

\begin{table}[]
\centering
\caption{MMS and selection percentages in Example 3.1. \label{tab:example1}}
\begin{tabular}{cc|cccccc|ccccc} \hline\hline
\multicolumn{13}{c}{$\varepsilon \sim N(0,1)$} \\ \hline
\multirow{2}{*}{$\bi{p}$}&\multirow{2}{*}{}&\multicolumn{6}{c|}{$K = 3\, (n = 60) $}&\multicolumn{5}{c}{$K=10\, (n = 200)$}\\  \cline{3-13}
&&MMS& RSD & $P_1$ & $P_2$ &$P_3$ &$P_{all}$ & MMS& RSD & $P_{min}$ & $P_{max}$ &$P_{all}$   \\ \hline
  I &$R^2$ & 3 & 0 &1&1&1&1  & 10&0 &1&1&1\\
 & MV& 3 & 0 &1&1&1&1  & 10&0 &1&1&1\\
& AD& 3 & 0 &1&1&1&1  & 10&0 &1&1&1\\
& DT& 3 & 0 &1&1&1&1  & 10&0 &1&1&1\\
& GN& 3 & 0 &1&1&1&1  & 10&0 &1&1&1\\ \\

 II &$R^2$ & 3 & 0 &1&1&1&1  & 10&0 &1&1&1\\
& MV& 3 & 0 &1&1&1&1  & 10&0 &1&1&1\\
& AD& 3 & 0 &1&1&1&1  & 10&0 &1&1&1\\
& DT& 3 & 0 &1&1&1&1  & 10&0 &0.992&1&0.992\\
& GN& 3 & 0 &1&1&1&1  & 10&0 &1&1&1\\ \\

 III &$R^2$ & 3 & 0 &1&1&1&1  & 10&0 &0.984&1&0.984\\
& MV& 3 & 0.75 &0.998&1&1&0.998 & 65&77.6 &0.328&1&0.328\\
& AD& 3 & 0 &1&1&1&1  & 11&3.73 &0.924&1&0.924\\
& DT& 15.5 & 20.3 &0.524&1&1&0.524  & 641&579 &0.016&1&0.016\\
& GN& 3 & 0 &1&1&1&1  & 10&2.98 &0.924&1&0.924\\  \hline
\multicolumn{13}{c}{$\varepsilon \sim t(2)$} \\ \hline

  I &$R^2$ & 5 & 45.5 &0.834& 0.876& 0.886&0.626 & 286&473 &0.136&0.848&0.136\\
& MV& 3 & 0 &1&0.998&0.998&0.996  & 10&0.75 &0.994&1&0.994\\
& AD& 3 & 0 &1&0.998&0.998&0.996  & 10&0 &0.996&1&0.996\\
& DT& 3 & 0 &0.996&0.994&0.994&0.988  & 10&0 &0.944&0.998&0.944\\
& GN& 3 & 0 &0.998&0.998&0.996&0.994  & 10&0 &0.994&1&0.994\\  \\

 II &$R^2$ & 5 & 46.6 &0.828& 0.880& 0.870& 0.634 & 314&623 &0.130&0.912&0.130\\
& MV& 3 & 0 &0.990&1&1&0.990  & 12&6.16 &0.876&1&0.876\\
& AD& 3 & 0 &0.994&1&1&0.994 & 10&2.24 &0.934&1&0.934\\
& DT& 3 & 0 &0.966&0.988&0.952&0.952 & 49&90.5&0.416&1&0.416\\
& GN& 3 & 0 &0.996&0.998&0.998&0.998  & 10&1.49 &0.950&1&0.950\\ \\

 III &$R^2$ & 26& 128 &0.564 &0.828& 0.880& 0.428  & 641&661 &0.008&0.914&0.008\\
& MV& 5 & 10.4 &0.726&1&1&0.726 & 174&211 &0.092&1&0.092\\
& AD& 4 & 5.22 &0.816&0.996&1&0.812 & 88&158 &0.286&1&0.286\\
& DT& 105 & 164 &0.090&0.990&0.996&0.088  & 999&617 &0.000&1&0.000\\
& GN& 3 & 2.98 &0.838&0.996&1&0.838 & 50&134 &0.444&1&0.444\\  \hline\hline
\end{tabular}
  \end{table}
  
From Table 1, we can see that  under the normal error term, each screening method works well when sample sizes are balanced or just slightly unbalanced. Heavy imbalance  affects the DT significantly.  One can see that the percentage of  identifying the active predictor $X_1$ in DT is reduced to 52.4\% while the minimal model size  increases from actual $3$ to $15.5$ for $K=3$. For 10 categorical LDA problem, heavy imbalance also affects performance of MV.  These effects are getting worse under the heavy-tailed error term. 
As expected, the $R^2$ method is not robust and it performs poorly under the heavy-tailed $t_2$ error. No methods perform well when $K=10$ with heavy-tailed error but GN has the very close estimation and highest probabilities  to include individual/all active predictors. 
GN performs the best and it is superior to others in this Example. \\

\noindent{\bf\textit{Example 3.2 (Binary Logistic Regression Model)}}

In this example, we consider a logistic regression model by following
the similar settings in He,  Ma and Xu \cite{He19} and Mai and Zou \cite{Mai13}. The binary response is generated by a logistic regression model as below. 
\begin{align*}
\log\left\{ \frac{P(Y=1|\bi X)}{P(Y=-1|\bi X)}\right\} = -3 +2 X_1+2X_2+2X_3+3\sin(X_4) +4X_5^2,
\end{align*}
where $\bi X \sim N(\bi 0, \bi \Sigma)$ with $\bi \Sigma =(\rho_{ij})_{p \times p}$ having two scenarios $\rho_{ij} =0$ and  $\rho_{ij} =0.5^{|j-i|}, i \neq j$.  The number of active predictors is five.  Two sample sizes $n=60$ and $n=200$ are considered. 
 
  \begin{table}[]
\centering
\caption{MMS and selection percentages in Example 3.2. \label{tab:example2}}
\begin{tabular}{ccc|cccccccc} \hline\hline
$n$ & $\Sigma_{ij}$ & Method & MMS &RSD & $P_1$ &  $P_2$ &$P_3$ &$P_4$ &$P_5$ &$P_{all}$ \\ \hline 
60 & $0$ &$R^2$ &  1178& 633 & 0.268 &0.310& 0.308& 0.270& 0.008& 0.000\\
&&  MV    &   571 & 558&0.250 &0.268 &0.290& 0.302 &0.064 &0.000 \\
&&  AD     &  554& 553 &0.260 &0.278 &0.292 &0.300 &0.192 &0.000\\
&& DT    &  716 & 612&0.210 &0.230 &0.230 &0.272 &0.178 &0.002\\
&& GN  & 556 & 543 & 0.248 &0.286 &0.292 & 0.298 &0.188 &0.000\\ \\ 
& $0.5^{|j-i|}$ & $R^2$&     501& 719&0.732& 0.920 &0.902 &0.636& 0.044& 0.024 \\
&&MV       & 130& 157& 0.674& 0.888 &0.878& 0.668 &0.136& 0.056\\
&&AD        &83 &109&0.692 &0.902 &0.884& 0.664& 0.194 &0.076\\
&&DT      &66 &113&0.608 &0.850 &0.854 &0.630 &0.536& 0.144\\
&&GN      &72 &  120&0.700 &0.900& 0.894& 0.664& 0.288& 0.130\\ \\ \hline

200 & $0$ &$R^2$ &  1034& 719 & 0.948 &0.952& 0.944& 0.906 &0.024&  0.018\\
&&  MV    &   14 & 23.1&0.918 &0.922 &0.926 &0.950& 1.000 &0.738 \\
&&  AD     &  12 & 21.8 &0.930 &0.930& 0.930& 0.944& 1.000 & 0.754 \\
&& DT    &  19 & 42.0 &0.882 &0.898& 0.898 &0.938 &1.000 & 0.656\\
&& GN  & 13  & 22.4 & 0.924& 0.932& 0.928& 0.954 &1.000 &0.756 \\ \\
& $0.5^{|j-i|}$ & $R^2$&     122& 277&1&  1&  1&  1 &0.308 &0.308 \\
&&MV       & 6&   2.24&   1&  1&  1&  1& 0.978& 0.978\\
&&AD        &5  & 0.00 & 1&  1 & 1 & 1& 1.000 &1.000\\
&&DT      & 5 &  0.00  & 1&  1 & 1 & 1& 1.000 &1.000\\
&&GN     &5  & 0.75  &   1&  1  &1 & 1& 1.000 &1.000\\ \hline\hline
\end{tabular}
 \end{table}

 From Table \ref{tab:example2}, it can be observed that the dependence structure among predictors helps identifying active predictors, and sample size $n=60$ is not sufficient. $R^2$ method is inferior in detecting $X_5$ which is of a quadratic form in the logit of response variable $Y$. The proposed GN method and AD method are among the best having the highest $P_{\textrm{all}}$ and the smallest MMS. \\

\noindent{\bf\textit{Example 3.3 (Genome-Wide Association Studies)}}

In this example, we consider the ultrahigh dimensional genome wide association studies (GWAS) by  following
the same setting in Cui,  Li and Zhong \cite{Cui15}. In GWAS, researchers allow to collect genetic data which
usually contain an extremely large number of single-nucleotide polymorphisms (SNPs).   The SNPs as predictors, in general,  are categorical with three classes, denoted by ${AA,Aa, aa}$. In this example, $Z_{ij}$ is denoted as the indictor of dominant effect of the $j^{th}$ SNP for $i^{th}$ subject and is generated in the following way
 \begin{align*}
Z_{ij} = \left \{ \begin{array} {rl} 
                       1, & \mbox{ if } \;\;X_{ij} < q_1 \\
                       0, & \mbox{ if }\;\; q_1 \leq X_{ij} < q_3,\\
                       -1,  & \mbox{ if }\;\; X_{ij} \geq  q_3
                       \end{array} \right.
\end{align*}
where $\bi X_i = (X_{i1},X_{i2},...,X_{ip}) \sim N(\bi 0, \bi\Sigma)$ for $i=1,2,..., n$ and $\bi\Sigma = (\rho_{jl})_{p \times p}$ with  $\rho_{jl} = 0.5^{|j-l|}$ for $j, l = 1,2,...,p$, and $q_1$ and $q_3$ are first and third quartiles of a standard normal distribution, respectively. The response is then generated by 
\begin{align*}
Y = \beta_1 Z_1 + \beta_2 Z_2+ 2 \beta_3 Z_{10} + 2 \beta_4 Z_{20} -2 \beta_5 |Z_{100}| + \epsilon,
\end{align*}
where $\beta_j = (-1)^U (2 \log (n)/\sqrt{n} +|Z_j|)$  for $ j=1,2,...,5$ with $U \sim $ Bernoulli(0.4) and $Z_j$ are iid from $N(0,1)$. The error term $\epsilon$ is from $N(0,1)$  or $t(1)$. There are five active SNP's $Z_1, Z_2, Z_{10}, Z_{20}$ and $Z_{100}$. We set $n = 200$ and $p = 2000$. The procedure is repeated 500 times. The results are reported in Table \ref{tab:example3}. 

\begin{table}[]
\centering
\caption{MMS and selection percentages in Example 3.3. \label{tab:example3}}
\begin{tabular}{cc|cccccccc} \hline\hline
 $\epsilon$ & Method & MMS &RSD & $P_1$ &  $P_2$ &$P_{10}$ &$P_{20}$ &$P_{100}$ &$P_{all}$ \\ \hline 
 $N(0,1)$ &$R^2$ &  8.0  &20.9& 0.938& 0.934 &0.996 &0.984& 0.880& 0.766\\
&  MV    &   12.5 & 42.9&0.904 &0.908& 0.982& 0.976& 0.846& 0.678 \\
&  AD     &  10.0  &32.1&0.922& 0.920 &0.986 &0.982& 0.854 &0.712\\
& DT    &  35.5 &127.8&0.790 &0.760& 0.940& 0.950& 0.854& 0.506\\
& GN  & 11.0 & 36.7 & 0.918 &0.914& 0.982& 0.980& 0.854& 0.698\\ \\ 
$t(1)$ & $R^2$&    1451.0 &851.8&0.284 &0.268 &0.388 &0.410& 0.278& 0.036 \\
&MV       & 50.5& 146.3& 0.834& 0.810 &0.948& 0.928 &0.774 &0.458\\
&AD        &51.0 &142.5&0.818 &0.814 &0.944 &0.928 &0.766& 0.446\\
&DT      &358.5 &584.9&0.506 &0.478& 0.720 &0.716& 0.652 &0.116\\
&GN      &64.0 &175.0&0.796  &0.788& 0.922 &0.894 &0.760 &0.412\\  \hline\hline
\end{tabular}
\end{table}
From Table \ref{tab:example3}, we can conclude that $R^2$ performs the best under the normal error, but the worst under the $t_1$ error term.  The DT method is not robust to the heavy-tailed errors.   It seems that DT has trouble in identifying weak signal predictors $Z_1$ and $Z_2$, in which $P_1$ and $P_2$ are quite low comparing to others. Our GN method performs similarly to AD and MV methods to effectively identify active categorical predictors which are linearly or nonlinearly correlated with the response. \\\\

\noindent{\bf\textit{Example 3.4 (Group predictors screening)}}

We compare performance of group screening methods in classification. Predictors $\bi X $ are grouped to be $ (\bi X_{(1)}, \bi X_{(2)},...,\bi X_{(r)})$, where $\bi X_{(l)} = (X_{3(l-1)+1},  X_{3(l-1)+2}, X_{3l})^T$ for $ l=1,...,r$.  In this example, there are two active predictor groups $\bi X_{(1)} = (X_1, X_2, X_3)^T$ and $\bi X_{(2)} = (X_4, X_5, X_6)^T$.  Again MMS and RSD are median and robust standard deviation of model sizes of the model including both groups. $P_1$, $P_2$ and $P_{\textrm{all}}$ are proportions including $\bi X_{(1)}$, $\bi X_{(2)}$ and both, respectively, in a model of the given size $d= [ n/\log n]$.

\begin{table}[]
\centering
\caption{MMS and selection percentages in Example 3.4. \label{tab:example4}}
\begin{tabular}{ccc|ccccc|ccccc} \hline\hline
&&&\multicolumn{5}{c|}{$\varepsilon \sim N(0,1)$}&\multicolumn{5}{c}{$\varepsilon \sim t(2)$} \\ \cline{4-13} 
Case & $\bi p$ & Method & MMS &RSD & $P_1$ &$P_2$ &$P_{all}$ & MMS &RSD & $P_1$ &$P_2$ &$ P_{all}$ \\ \hline
(a) & I & EN & 27 & 57.1 & 0.634 &0.636 &0.392& 39.5 & 86.7&0.566 &0.554 &0.304\\
&& ENR& 18 & 43.3  &0.662& 0.690& 0.462&19  &48.7& 0.658 &0.694 &0.452\\
&&GI &712 & 391& 0.112 &0.104 &0.010& 620& 387&0.116 &0.108 &0.008\\
&& DT & 2 & 0& 1&1&1 & 16 & 47.0& 0.692& 0.696 &0.486\\
&& GN &  2 & 0& 1&1&1 & 2 & 0.75 &0.968 &0.962 &0.930\\\\

& II &EN  & 133&  209& 0.310& 0.330 &0.114&159& 248.9&0.250& 0.298 &0.076\\
&& ENR &  22&  49.4&  0.630 &0.640& 0.406&32& 74.3& 0.566 &0.610 &0.336\\
&&GI &  912&173&0.022& 0.030& 0.000& 882& 257& 0.024 &0.034& 0.000\\
&& DT& 2 & 0& 1&1&1 & 20 &50.9 & 0.678 &0.686 &0.430 \\
& & GN & 2 & 0& 1&1&1 & 2  &1.50& 0.952 &0.950 &0.904\\ \\

&III&EN  & 640& 325.4& 0.046& 0.040& 0.002&591& 358& 0.058 &0.050 &0.004\\
&& ENR   & 7 & 18.7&  0.796 &0.796 &0.654&20 & 44.8& 0.658 &0.644& 0.426\\
&&GI  & 998&  11.9& 0.000 &0.000& 0.000&991& 33.6& 0.002 &0.000 &0.000\\
&& DT & 2 & 0& 1&1&1 & 25 & 59.7&0.600 &0.610 &0.378\\
& & GN & 2 & 0& 1&1&1 & 3 &  4.48& 0.908& 0.906 &0.830 \\ \\

(b) & I & EN  &  64 &166& 0.872& 0.288 &0.240&94 &192&0.726& 0.244& 0.168\\
&&ENR &  86 &150& 0.926& 0.218& 0.200&93.5& 188& 0.850& 0.222& 0.192\\
&&GI &  698 &383&0.142 &0.048 &0.006&671& 386&0.120 &0.056 &0.012\\
&&DT &  2 & 0 &1 &0.990& 0.990& 62 &134 &0.890 &0.238& 0.212\\
&& GN &  2 & 0 & 1& 0.990& 0.990& 7  &23.9&1.000 &0.632& 0.632\\\\

& II & EN  & 195& 272& 0.514& 0.130 &0.064&230&277&0.426 &0.102 &0.048\\
&& ENR&77 &154& 0.924 &0.244 &0.222&94 &177& 0.824 &0.202& 0.168\\
&&GI  & 912&169& 0.034 &0.018 &0.002&849& 238&0.036 &0.024 &0.000\\
&&DT & 2 &0& 1 &0.994& 0.994&57.5 &120&0.926 &0.236& 0.214 \\
&& GN &2 &0&1& 0.994& 0.994& 6 & 23.5&  0.998& 0.644 &0.642 \\ \\

& III&EN&623& 338& 0.048& 0.026& 0.000&592& 324&0.048& 0.028& 0.000\\
&&ENR& 52& 118& 0.986& 0.290 &0.286&66.5& 153&  0.882 &0.296& 0.254\\
&&GI&1000&  4.48& 0.000 &0.002& 0.000&993& 25.7& 0.002 &0.004 &0.000\\
&& DT &  2  &    0.75&  1 &0.910&0.910&  76.5 &144& 0.900 &0.198 &0.176 \\
&& GN & 2    &  0  &1 &0.970& 0.970& 17  &51.9& 0.990 &0.476 &0.468 \\  \hline\hline 
\end{tabular}
\end{table}

We consider $K=3$.  For each $i^{th}$ observation, the categorical response $Y_i$ is generated from three discrete distributions with 3 categories with $\bi p = (p_1, p_2,p_3)$ with $p_k = P(Y_i = L_k)$: (I) balanced, $\bi p =(1/3,1/3,1/3)$; (II) slightly unbalanced, $\bi p = (3/12, 4/12, 5/12)$ and (III) heavily unbalanced, $\bi p = (0.1,0.3,0.6)$. We consider two cases. In Case (a),  two groups have a same role to predict categorical variable $Y$. In Case (b), $\bi X_{(2)}$ is a weaker signal than $\bi X_{(1)}$. More specifically,  given $Y_i=L_k$, the $i^{th}$  predictors $\bi X_ i$ are generated by the location model $\bi X_i =\bi \mu_G^{(k)} +\bi \epsilon_i$, where $\bi \epsilon_i$ is a $d$-dimensional error term independently from $N(0,1)$ or $t(2)$ and $\bi \mu_G^{(k)} = (\bi \mu_1^{(k)}, \bi \mu_2^{(k)}, 0, ...,0)^T$. Let $\bi \mu^{(1)} =(1,0,0)^T, \bi \mu^{(2)} =(0,1,0)^T$, and $\bi \mu^{(3)} =(0,0,1)^T$. In Case (a), 
$\bi \mu_1^{(k)} =\bi \mu_2^{(k)} =1.5 \bi \mu^{(k)}$. In Case (b), $\bi \mu_1^{(k)} =2  \bi \mu^{(k)}$, while $\bi \mu_2^{(k)} =\bi  \mu^{(k)}$. We set sample size $n$ to be 200 and group number $r$ to be 1000. The repetition number $M$ is $500$.

Since EN, ENR and GI are designed for categorical group screening, each component of $\bi X_i$ has to be discretized and treated as a categorical variable. As suggested by authors who proposed the methods,  the standard normal quantiles $(-0.67, 0, 0.67)$ are used for slicing each variable into 4 categories.

From Table \ref{tab:example4}, one can see that EN, ENR and GI completely fail in all cases, which is not surprising because they lose too much information in the slicing process. Also, their computation burden is prohibitive with computational complexity exponentially increasing on group size and polynomially increasing on slice numbers.  Under normal error term, the DT and GN work perfectly in case (a) and GN has a slight edge over DT in Case (b) III with slightly higher $P_2$ and $P_{\textrm{all}}$ percentages. While under $t(2)$ error term, the advantage of GN over DT becomes noticeable with much smaller MMS and higher identification percentages. 

\section{Real data analysis}\label{sec:realdata}
We compare the screening methods in gene selection and classification on two cancer data sets. One is the human lung carcinomas data that have been studied by Bhattacharjee et al.
\cite{Bhattacharjee01} and Cui, Li and Zhang \cite{Cui15}.  The second one is the TCGA breast cancer microarray dataset from the UCSC Xena database \cite{Goldman18}. 
\subsection{Lung cancer classification}

There are 203 samples on 12,600 mRNA expression levels in the Lung carcinomas data set. 
The 203 specimens are classified into five subclasses: 139 in lung adenocarcinomas
(ADEN), 21 in squamous cell lung carcinomas
(SQUA), 6 in small cell lung carcinomas (SCLC), 20 in pulmonary
carcinoid tumors (COID), and the remaining 17 normal
lung samples (NORMAL). The data have been standardized to zero mean and unit variance for each gene. 

Following the similar procedure in Cui, Li and Zhang \cite{Cui15}, we randomly partition approximately $100\alpha\%$ of the observations from each subclass as the training samples and the remaining observations as the testing samples. $\alpha$ is set to be 0.9 and 0.8. 
Each screening method with LDA are applied to the training set and their performance is evaluated by the testing samples. More specifically,  each screening method first reduces dimensionality to $d = \lfloor n/\log n\rfloor =  38$ based on the training data. Then one chooses the optimal model size from 2 to $d$ by the leave-one-out cross-validation in LDA. The procedure is repeated 100 times. The mean and standard deviation of training classification errors, test classification errors and model sizes are reported in Table \ref{tab:lungcancer}. Using LDA classification method, MV is the best in terms of the smallest test error and the smallest model size. GN has a second smallest test error, but its model size is relatively large. $R^2$ is the worst with the largest errors and the largest model sizes. We also can observe that for each screening method, the test error is much larger than the training error; the test error is more than twice of the training error. This suggests the overfitting problem of the classification method LDA. To overcome this limitation, the popular random forest (RF) is used as classification method to evaluate screening methods. 

Based on the training data, the first step again is to select $d_1$ genes by each screening method. The second step is further to reduce to the optimal size. Rather than using leave-one-out cross validation, we repeat 20 times of RF with 200 trees for each model of sizes from 2 to $d_1$. The model with the smallest average out of bag classification errors is selected to be the optimal. Then the training error and test error are obtained for the optimal model under 500 random trees. The procedure is repeated 100 times. The mean and standard deviation of  classification errors and model sizes are reported in Table \ref{tab:lungcancer}. 

\begin{table}[] 
\centering
\caption{Errors in $\%$ and model size selected by each screening method with classification methods of LDA and random forest. Standard deviations are in parentheses. \label{tab:lungcancer}}
\begin{tabular}{cc|rrr|rrr} \hline
&& \multicolumn{3}{c|}{LDA} & \multicolumn{3}{c}{Random Forest} \\ \cline{3-8}
$\alpha$& Method& Train err & Test err & Size  & Train err & Test err&Size \\ \hline
0.9 & R2 &  6.76(1.79) & 14.33(6.58) & 33.5(4.26) & 11.35(1.45)& 12.43(4.88) & 34.7(3.61) \\
& MV& 3.44(1.02)&8.67(5.92) &20.8(6.10)& 5.74(0.68) &  8.20(5.44) & 25.4(7.60) \\
& AD & 4.25(1.41) &9.10(5.79) &24.7(9.10) &6.05(0.69) &8.86(5.58) &20.5(8.21) \\
& DT & 4.80(1.86)&10.57(5.85) &29.2(7.44) & 9.56(1.25) &11.14(5.63)& 20.9(10.4) \\
& GN & 4.59(1.11)&8.95(5.76) &31.3(5.15)&6.35(1.14) &7.98(4.53) &30.9(4.62) \\  \\
0.8 & R2 & 6.45(2.08) &14.8(4.35)&31.7(5.39) &11.38(1.86)& 11.90(3.28)&34.9(3.91) \\
& MV& 3.55(1.41)&7.98(4.06) & 20.6(7.09) &5.01(0.84) &6.88(3.73) &25.3(8.82) \\
&AD & 4.29(1.60)&8.88(4.57)&22.6(9.09)&6.12(0.82) &7.55(3.99) &20.7(7.53) \\
& DT & 4.09(1.95) &9.83(4.34)&29.8(6.45) &9.70(1.43) &10.78(3.44) &20.0(8.15) \\
& GN &3.91(1.72) & 8.78(4.41)&31.0(6.45)& 6.50(1.23) &7.15(3.36) & 30.6(5.34)\\ \hline
\end{tabular}
\end{table}

The differences between testing error and training error of each screening method in RF are much smaller than those in LDA.  The RF yields smaller testing errors than those of the LDA for all screening methods except for DT. The reason that DT has a larger testing error in RF than it in LDA is probably due to its small model size selected by RF. For the GN screening method, LDA and RF select similar model size, but the test classification in RF is smaller than it in LDA. The classification improvement of the AD is even built on a smaller model in RF. Overall, GN performs well, sometimes the best, but its model size is relatively large. 

\subsection{Breast Cancer Gene Selection}
The TCGA breast cancer microarray dataset from the UCSC Xena
database \cite{Goldman18} contains expression levels of 17278 genes from 506 patients. 
Patients are classified into four subtypes:  luminal A, luminal B, HER2-enriched, and basallike.
For breast cancer, PAM50 is a gene signature consisting of 50 genes and is
considered as the gold-standard for breast cancer subtype prognosis and prediction \cite{Parker09}. This data set contains 48 genes of PAM50.  

We demonstrate two group screening methods DT and GN in selecting the signature gene group called PAM48. 

We consider four types of comparison groups in the remaining 17230 genes. The first type is to form groups in the original order of genes in the dataset. The group sizes are considered to be 50,100, 150 and 200.  The corresponding numbers of groups $r$ are 344, 172,  114, and 86, respectively. The second type is to form random groups with gene indices sampling from 1 to 17230  without replacement, while the third type is for random groups with replacement. The fourth type is to obtain groups by the ordering of marginal correlations of each gene with cancer subtype categories.  For each type, the group with the highest correlation is selected for comparison of PAM48.  The models of the selected genes by those group types are denoted as Original, Random1, Random2 and Marginal, respectively. 

Also we compare classification performance of the models with the selected group of genes. We randomly hold-out 20\% as test data. Based on the remaining 80\% training data, we select the top group by the Gini or distance method. We use random forest (RF) method for classification.  We repeat 10 times of RF with 200 trees for each model,  then evaluate performance of the selected model by training and test classification errors. 

\begin{figure}[]
 \center
\begin{tabular}{cc}
\includegraphics[width=8cm, height=8cm]{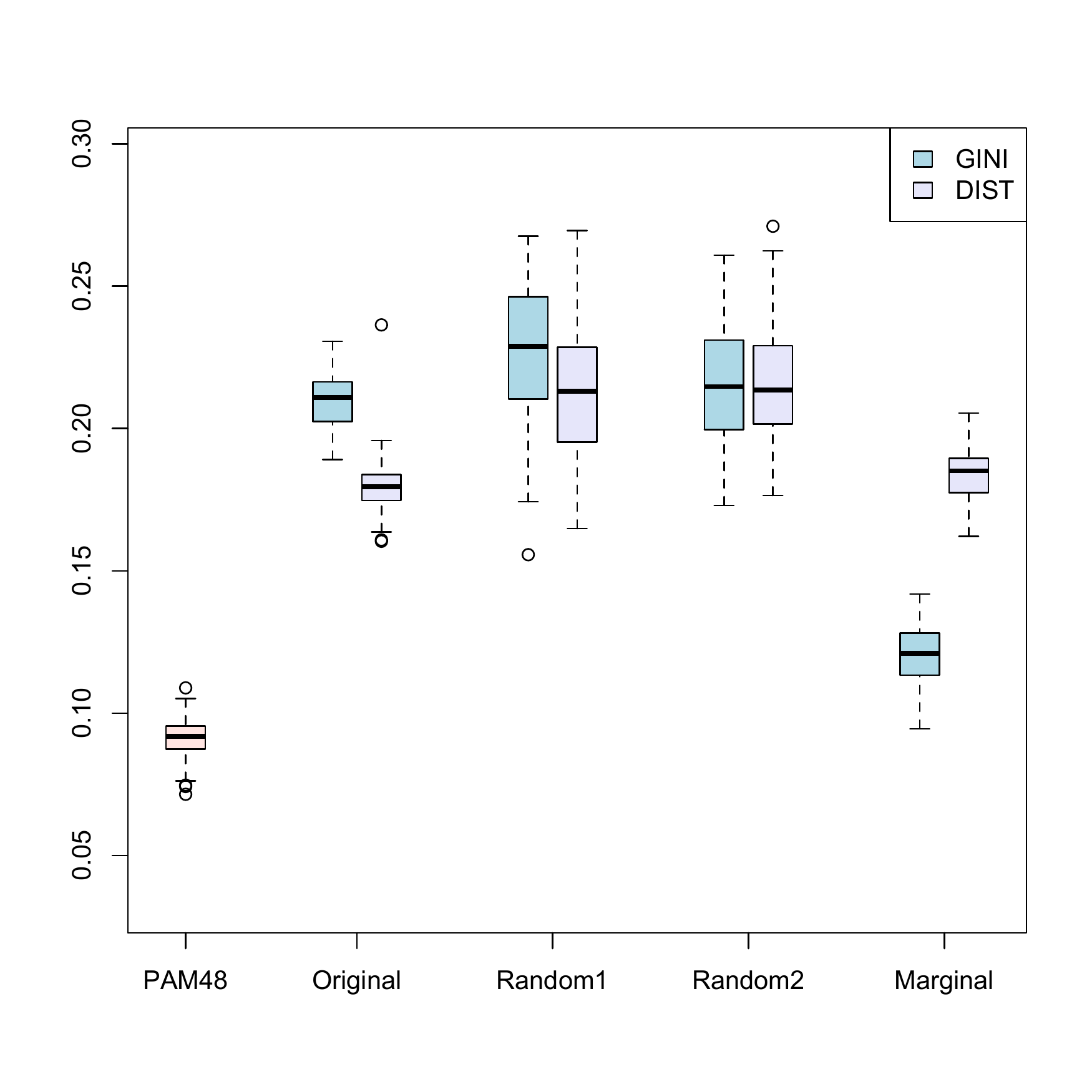}&
\includegraphics[width=8cm, height=8cm]{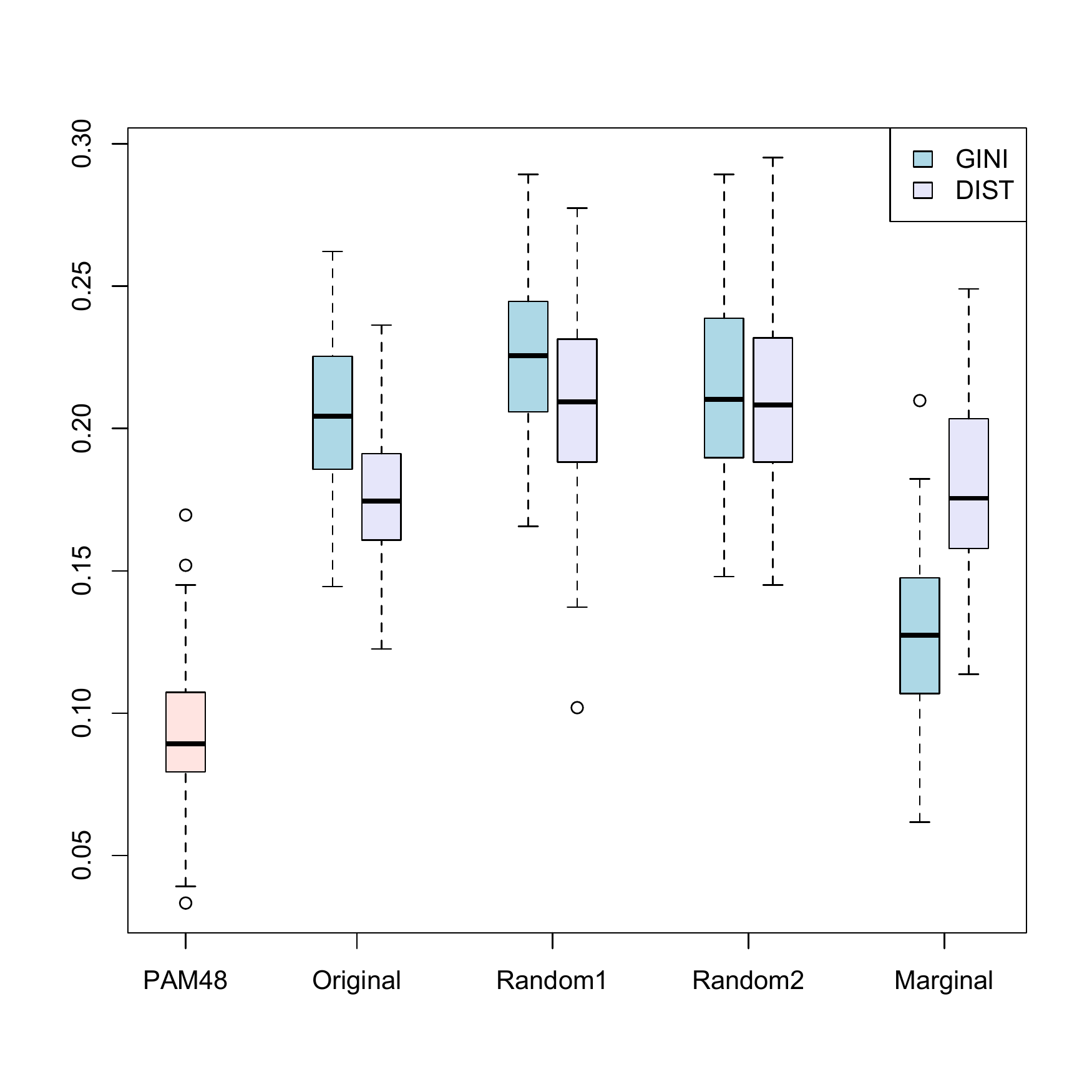}
\vspace{-0.15in} \\ 
 {\small (a) Training Misclassification Rate} & {\small (b) Testing Misclassification Rate}
\end{tabular}
\caption{Group screening comparison of the Gini and distance methods. (a) Boxplot of misclassification training errors for the selected models of size 50 comparing with the PAM48 model.  (b) Boxplot of misclassification testing errors for the selected models of size 50 comparing with the PAM48 model.  \label{fig:BC}}
\end{figure}

The whole procedure is repeated 100 times. We summarize the means with standard deviations of the training and test classification errors as well as correlations of the selected groups and PAM48  in Table \ref{tab:BC}. Their distributions of classification errors for the models with group size of 50 are displayed in Figure \ref{fig:BC}. 

\begin{table}[] 
\centering
\caption{Correlations and RF classification errors in $\%$ of the selected models comparing with PAM48. Standard deviations are in parentheses. \label{tab:BC}}
\begin{tabular}{ll|cc|cc|cc} \hline\hline
Model & \multirow{2}{*}{Model} &  \multicolumn{2}{c|}{Group Correlations}&  \multicolumn{2}{c|}{Training Error} & \multicolumn{2}{c}{Testing Error}\\ \cline{3-8}
Size&& GN& DT & GN& DT & GN& DT \\ \hline
& PAM48 &.269(.004)&.750(.007)&\multicolumn{2}{c|}{9.128(.696)}&\multicolumn{2}{c}{9.280(2.531)}\\ \hline
50 & Original  &.186(.004)&.695(.008)&21.0(0.93)&18.0(0.89)&20.5(2.70)&17.6(2.31)\\
& Random1 &.147(.012)&.688(.012)&22.7(2.37)&21.4(2.32)&22.5(2.84)&21.0(3.07)\\
& Random2 &.153(.017)&.693(.011)&21.6(2.13)&21.6(2.07)&21.2(3.26)&21.0(3.15)\\
& Marginal &.260(.005)&.698(.006)&12.0(1.03)&18.4(0.90)&12.7(2.97)&18.1(3.00)\\ \\
100 & Original  &.133(.002)&.686(.008)&19.8(1.01)&19.3(1.48)&19.3(2.51)&18.6(3.21)\\
& Random1 &.118(.008)&.697(.011)&20.9(1.94)&20.0(2.14)&20.0(3.33)&19.0(2.71)\\
& Random2 &.124(.014)&.699(.012)&20.4(1.91)&20.0(1.88)&19.8(2.82)&19.5(2.95)\\
& Marginal &.245(.005)&.703(.006)&11.3(0.72)&14.9(0.98)&11.8(2.80)&15.0(2.71)\\ \\
150 & Original  &.112(.002)&.700(.007)&15.8(0.87)&15.8(0.71)&15.1(2.55)&15.0(2.39)\\
& Random1 &.104(.007)&.698(.010)&19.4(1.77)&19.5(1.89)&18.8(2.86)&18.5(3.07)\\
& Random2 &.108(.009)&.704(.011)&19.5(1.67)&19.3(1.69)&18.5(2.95)&18.6(2.74)\\
& Marginal &.228(.004)&.705(.005)&11.5(0.74)&13.2(0.92)&11.8(2.71)&13.4(2.96)\\ \\
200 & Original  &.103(.002)&.706(.007)&16.4(0.99)&15.8(0.70)&15.9(2.64)&15.1(2.36)\\
& Random1 &.095(.006)&.699(.010)&19.3(1.74)&19.1(1.74)&18.7(3.11)&18.2(2.68)\\
& Random2 &.099(.008)&.704(.012)&19.1(1.71)&18.8(1.53)&18.4(2.95)&18.0(2.93)\\
& Marginal &.215(.005)&.706(.006)&11.6(0.73)&12.4(0.74)&12.1(2.73)&12.5(2.89)\\ \hline\hline
\end{tabular}
\end{table}

First of all, PAM48 is ranked the top among all considered groups in terms of Gini correlation or distance correlation. In this sense, both group  screening methods successfully select the signature gene group.  From Table \ref{tab:BC}, we see that PAM48 has the mean Gini correlation 0.269, while the second highest correlated group is the Marginal group with the mean ranging from 0.260 to 0.215 for group sizes from 50 to 200. Two random groups have the lowest Gini correlations. Similar observations hold for the distance correlation.  However, the differences of distance correlations between the Marginal group and other groups are small. 

For classification performance, PAM48 works the best with the smallest training error and testing error. The Marginal group performs the second best. The pre-orderings really helps reduce misclassification.  Especially in Gini correlation ordering, the improvement is large with almost 8\% error decrease  in the model of size 50. Even in the model of size 150 or 200, a 3\% drop from that of Original group is visible. The distance correlation ordering benefits larger models, but does not aid the model of size 50. Two random groups preform inferior to others for all sizes. It is reasonable that testing errors have larger variation than training errors and two randomness models have bigger standard deviations in errors than the others. 

Those results for the models of size 50 can be viewed from Figure \ref{fig:BC}. Both plots are in the same scale and hence comparable. The boxplots in (b) have longer length of interquartile ranges and whiskers. It seems that the original ordering assists model selection in the DT method, but its improvement over random groups in the GN method is slight. Again, except for PAM48, the Marginal group model selected by the GN method perform much better than the others. 

\section{Conclusions and Discussion}\label{sec:conclusion}

In the literature, the sure screening procedures for marginal predictor are fruitful since Fan and Lv \cite{Fan08}. 
However, procedures for screening grouped features in ultrahigh-dimensional data are very limited. 
We have proposed a GDC based screening procedure, GN,  for both the individual and the grouped features when the response is categorical. The proposed one is easy to implement, model-free, and holds the sure screening property. Under a stronger separability condition,  the stronger theoretical result called the ranking consistency property has been established for the proposed screening procedure. 
 The simulation studies have shown that the proposed GN is superior to most of the current existing methods for both individual and grouped features screening.

Although our GN procedure is majorly for numerical covariates and the Gini impurity method in~\cite{Wang23} is for categorical covariates, it is interesting to see some connection between them. Define the discrete distance between two categorical variables as 
\begin{align*}
\| \bi x_1 -\bi x_2 \|= \mathbbm{1}(\bi x_1 \neq \bi x_2), 
\end{align*}
where $\mathbbm{1}(\bi x)$ is the indicator function.   Then the Gini mean difference of a categorical random variable is its Gini impurity,  the commonly used index for the assessing classification performance. That is, for response categorical variable $Y$, 
\[\E \|Y_1-Y_2\| = \sum_{k=1}^K P(Y_1 = L_k) (1-P(Y_2 \neq L_k)) = 1-\sum_{k=1}^K p_k^2 = \textrm{GI}(Y). \]
Further, if the group covariates $\bi X_{(l)}$ are categorical, our Gini correlation defined in (\ref{mgc}) turns to be
\begin{align}\label{ccgr}
\textrm{gCor}(\bi X_{(l)}, Y)=\rho_l = \frac{\textrm{GI}(\bi X_{(l)}) - \textrm{GI}(\bi X_{(l)}|Y)}{\textrm{GI}(\bi X_{(l)})},
\end{align}
while the index used in ~\cite{Wang23} is 
\begin{align*} 
\textrm{GI} (X_{(l)}, Y ) = \frac{\textrm{GI}(Y) -\textrm{GI}(Y|\bi X_{(l)})}{\log N_l},  
\end{align*}
where $N_l$ is the total number of levels of $\bi X_{(l)}$. Intuitively, our standardized term in (\ref{ccgr}) is more natural and takes more information of $\bi X_{(l)}$. Hence, when group covariates are categorical, the proposed screening method is ready to apply.

\section{Appendix}
We first give lemmas that will be used to prove the theorems.
Lemmas \ref{lemma1} and \ref{lemma2} are from \cite{Serfling80} and  \cite{Li12}, and Lemma \ref{kk} is from Lemma A.3. in \cite{Cui15} derived from Bernstein's Inequality.    

\begin{lemma}\label{lemma1}
Let $W$ be a random variable with $P(a \leq W \leq b) =1$ and $\mu = \E W$, then
\[
\E [\exp\{s(W-\mu)\}] \leq \exp\{s^2(b-a)^2/8\},\,\,\,\mbox{ for any } s>0.
\]
\end{lemma}
\begin{lemma}\label{lemma2}
Let $h(X_1,X_2,...,X_m)$ be a kernel of the $U$-statistic $U_n$, and $\theta = \E h(X_1,...,X_m)$. If $a\leq h(X_1,X_2,...,X_m)\leq b$, then 
\[
Pr(U_n -\theta \geq t) \leq \exp\{-2[n/m] t^2/(b-a)^2\}, \,\,\, \mbox{ for any } t>0, \mbox{ and } n\geq m,
\]
where $[n/m]$ denotes the integer part of $n/m$. 
\end{lemma}
\begin{lemma}\label{kk}
For $k=1, .., K$, let $p_k=F(Y=L_k)$ and $\hat{p}_k=\dfrac{1}{n}\sum_{i=1}^n I(Y=L_k)$, we have 
$$P(\abs{\hat{p}_k-p_k}\geq \varepsilon) \leq 2 \exp\{-\dfrac{n\varepsilon^2}{2(p_k+\varepsilon/3)}\}$$
for any $\varepsilon \in (0, 1)$.
\end{lemma}

We prove the inequality (\ref{uni_con}) by showing the uniform consistency of the denominator and the numerator of $\hat{\rho}_l$, respectively. 

For $l =1,...,r$, the  Gini covariance between $\bi X_{(l)}$ and $Y$ in (\ref{Gcov})
can be estimated by 
\begin{align*}
\hat{\mbox{gCov}}(\bi X_{(l)}, Y)&={n \choose 2}^{-1}\sum_{1 \leq i <j \leq n}\|\bi X_{(l)i}-\bi X_{(l)j}\|-\sum_{k=1}^K \hat{p}_k {n_k \choose 2}^{-1}\sum_{1 \leq i <j \leq n_k}\|\bi X_{(l|k)i}-\bi X_{(l|k)j}\|\\
&:=U_{n,l}-\sum_{k=1}^K \hat{p}_k U_{n_k, l}.
\end{align*}

Let $h(\bi X_{(l)1}, \bi X_{(l)2})=\|\bi X_{(l)1}-\bi X_{(l)2}|$ be the kernel to $U_{n,l}$, and it can be  decomposed as: $h=h   \mathbbm{1}(h>M)+h   \mathbbm{1}(h\leq M)$, where $M$ will be provided later. Then
\begin{align*}
U_{n,l}&={n \choose 2}^{-1}\sum_{1 \leq i <j \leq n}h(\bi X_{(l)i}, \bi X_{(l)j})\mathbbm{1}(h\leq M)+{n \choose 2}^{-1}\sum_{1 \leq i <j \leq n}h(\bi X_{(l)i}, \bi X_{(l)j})\mathbbm{1}(h> M)\\
&:=\hat{R}_{l,1}+\hat{R}_{l,2}
\end{align*}
and
\begin{align*}
&\mathbb{E}\|\bi X_{(l)1}-\bi X_{(l)2}\|\\
&=\mathbb{E}\left\{\|\bi X_{(l)1}-\bi X_{(l)2}\|\mathbbm{1}(h(\bi X_{(l)1}, \bi X_{(l)2})\leq M)\right\}+\mathbb{E}\left\{\|\bi X_{(l)1}-\bi X_{(l)2}\|\mathbbm{1}(h(\bi X_{(l)1}, \bi X_{(l)2})> M)\right\}\\
&:=R_{l,1}+R_{l,2}.
\end{align*}
We have $\mathbb{E}\hat{R}_{l,1}=R_{l,1}$ and $\mathbb{E}\hat{R}_{l,2}=R_{l,2}$. Applying Markov's inequality, we obtain that for any $\varepsilon>0$ and $\gamma>0$
\begin{align*}
P(\hat{R}_{l,1}-R_{l,1} \geq \varepsilon)\leq \exp(-\gamma \varepsilon) \exp(-\gamma R_{l,1}) \mathbb{E}\{\exp(\gamma \hat{R}_{l,1})\}.
\end{align*}
By the fact that any $U$ statistic can be represented as an average of averages of i.i.d random variables in \cite{Serfling80}, we can have 
$$\hat{R}_{l,1}=(n!)^{-1} \sum_{n!}\Lambda_1(\bi X_{(l)i_1}, \bi X_{(l)i_2},..., \bi X_{(l)i_n}),$$ where  $\sum_{n!}$ denotes summation over all $n!$ permutations $(i_1,...i_n)$ of $(1,...,n)$ and 
each $\Lambda_1$ is an average of $m=[n/2]$ i.i.d random variables. More specifically, 
\begin{align*}
{\Lambda_1(\bi x_1,...,\bi x_n)=\dfrac{1}{m}\sum_{t=1}^m h( \bi x_{2t-1}, \bi x_{2t}) \mathbbm{1}(h( \bi x_{2t-1}, \bi x_{2t})\leq M).}
\end{align*}
Following Jensen's inequality, for $1 < \gamma \leq 2s_0$,
\begin{align*}
\mathbb{E}\left \{ \exp(\gamma \hat{R}_{l,1}) \right \}&=\mathbb{E}\left \{  \exp \bigg( \gamma (n!)^{-1} \sum_{n!}\Lambda_1(\bi X_{(l)1}, \bi X_{(l)2},..., \bi X_{(l)n})\bigg)\right \}\\
& \leq (n!)^{-1} \sum_{n!} \mathbb{E} \left\{  \exp \bigg( \gamma \Lambda_1(\bi X_{(l)1}, \bi X_{(l)2},..., \bi X_{(l)n})\bigg)
\right \}\\
&=\mathbb{E}^m \left \{ \exp\bigg(\dfrac{1}{m} \gamma h(\bi X_{(l)1}, \bi X_{(l)2})\mathbbm{1}(h(\bi X_{(l)1}, \bi X_{(l)2})\leq M)  \bigg)\right\},
\end{align*}
which, together with Lemma 1, entails immediately that 
\begin{align*}
P(\hat{R}_{l, 1}-R_{l,1} \geq \varepsilon) &\leq \exp(-\gamma \varepsilon) \exp(-\gamma R_1) \mathbb{E}\{\exp(\gamma \hat{R}_{l,1})\}\\
& \leq  \exp(-\gamma \varepsilon) \exp(-\gamma R_{l,1}) \mathbb{E}^m \left \{ \exp\bigg(\dfrac{1}{m} \gamma h^{(t)} \mathbbm{1}(h^{(t)}\leq M)  \bigg)\right\}\\
&=\exp(-\gamma \varepsilon) \mathbb{E}^m \left \{ \exp \bigg(\dfrac{1}{m} \gamma [h^{(t)} \mathbbm{1}(h^{(t)}\leq M) -R_{l,1}] \bigg)\right\}\\
& \leq \exp\bigg(-\gamma \varepsilon +M^2 \gamma^2/(8m) \bigg).
\end{align*}
Choose $\gamma=4 \varepsilon m/M^2$ and we have $P(\hat{R}_{1, l}-R_{l,1} \geq \varepsilon)\leq \exp (-2 \varepsilon^2 m/M^2)$. 
Therefore, 
\begin{align}\label{R1}
P(\abs{\hat{R}_{l, 1}-R_{l,1}} \geq \varepsilon)\leq 2 \exp (-2 \varepsilon^2 m/M^2).
\end{align}

Next we show the consistency of $\hat{R}_{l,2}$. With Cauchy-Schwartz and Markov's inequalities,
\begin{align*}
R^2_{l,2}& \leq \E\big[h^2(\bi X_{(l)i}, \bi X_{(l)j})\big] P\big(h(\bi X_{(l)i}, \bi X_{(l)j})>M\big)\\
& \leq \E\big[h^2(\bi X_{(l)i}, \bi X_{(l)j})\big]   \E \big[\exp{\big(s_1h(\bi X_{(l)i}, \bi X_{(l)j})\big)}\big]/\exp{(s_1M)}\\
&\leq 2\E\big[\|\bi X_{(l)i}\|^2\big]   \{\E \big[\exp{(s_1\|\bi X_{(l)i}\|)}\big]\}^2  /\exp{(s_1M)}
\end{align*}
for any $s_1>0$.
%
  
  If we choose $M=cn^{\beta}$ for $0<\beta<1/2-\kappa$, with condition \textbf{C2},  $R_{l,2}\leq  \varepsilon/2$ when $n$ is sufficiently large. Thus we have
  \begin{align*}
  P(|\hat{R}_{l,2}-R_{l,2}|> \varepsilon)\leq P(|\hat{R}_{l,2}|>\varepsilon/2).
  \end{align*}
  Furthermore,
  $\{|\hat{R}_{l,2}|>\varepsilon/2\} \subset \{\|\bi X_{(l)i}\|> M/2, \text{ for some $1\leq i \leq n$}\}$. 
  In fact, if $\|\bi X_{(l)i}\|\leq M/2$ for all $1\leq i \leq n$, then $h(\bi X_{(l)i}, \bi X_{(l)j}) \leq \|\bi X_{(l)i}\|+\|\bi X_{(l)j}\| \leq M$ and $\hat{R}_{l,2}=0$, which contradicts the fact that $\hat{R}_{l,2}>\varepsilon/2$.  
  By condition $\textbf{C2}$ and Markov's inequality, we have 
  \begin{align*}
  P(\|\bi X_{(l)i}\|> M/2) \leq C \exp{(-sM/2)}.
  \end{align*}
 And hence
 \begin{align*}
 \max_{1 \leq l \leq r}P(|\hat{R}_{l,2}|>\varepsilon/2)& \leq n \max_{1 \leq l \leq r}P(\|\bi X_{(l)i}\|> M/2)\\
 & \leq Cn \exp{(-sM/2)}
 \end{align*}
\begin{align}\label{R2}
P(\abs{\hat{R}_{2, l}-R_{l,2}} > \varepsilon) & \leq P(\abs{\hat{R}_{2,l}}>\varepsilon/2)\nonumber \\
& \leq  nC \exp(-s M/2).
\end{align}
Combining (\ref{R1}) and (\ref{R2}), 
\begin{align}\label{0-0}
P(\abs{U_{n, l}-\Delta_{l} }\geq 2 \varepsilon) \leq 2 \exp(-\varepsilon^2 n^{1-2\beta})+n C \exp(-s n^{\beta}/2),
\end{align}
where $0< \beta<1/2-\kappa$.
With the same arguments, for $k=1,..,K$, we have
\begin{align}\label{l-l}
P(\abs{U_{n_k, l}-\Delta_{l|k} }\geq 2 \varepsilon) \leq 2 \exp(-\varepsilon^2 n_k^{1-2\beta})+n_k C \exp(-s n_k^{\beta}/2).
\end{align}

With the results of (\ref{0-0}) and (\ref{l-l}), we are ready to prove the concentration result of sample Gini covariance which stated in the below Lemma. 

\begin{lemma}\label{lemma4} Under conditions \textbf{C1} and \textbf{C2}, for any $\varepsilon \in (0,1)$, $l=1,...,r$, and $0< \beta<1/2-\kappa$, we have
\begin{align*}
&P(\abs{\hat{\mbox{\textrm{gCov}}}(\bi X_{(l)}, Y)-\mbox{gCov}(\bi X_{(l)}, Y)}\geq \varepsilon) \\
&\leq 2K_n  \exp(-\varepsilon^2 n^{1-2\beta})+n K_n \exp(-s n^{\beta}/2)+ 2K_n \exp\{-C \dfrac{n}{K_n}\varepsilon^2\} .
\end{align*}
\end{lemma}
\noindent{\textbf{Proof of Lemma \ref{lemma4}.}}
\begin{align*}
\hat{\mbox{gCov}}(\bi X_{(l)}, Y)-\mbox{gCov}(\bi X_{(l)}, Y) &=U_{n,l}-\sum_{k=1}^K \hat{p}_k U_{n_k, l}-(\Delta_{l}-\sum_{k=1}^K p_k \Delta_{l|k})\\
&=U_{n,l}-\Delta_{l}-\sum_{k=1}^K \hat{p}_k U_{n_k, l}+\sum_{k=1}^K p_k \Delta_{l|k}\\
&=(U_{n,l}-\Delta_{l})-\sum_{k=1}^K \hat{p}_k (U_{n_k, l}- \Delta_{l|k})-\sum_{k=1}^K (\hat{p}_k-p_k) \Delta_{l|k}\\
&:=\hat{I}_{l,1}+\hat{I}_{l, 2}+\hat{I}_{l,3}
\end{align*}
By (\ref{0-0}), 
\begin{align}\label{I1}
P(\abs{\hat{I}_{l,1} }\geq  \varepsilon) \leq 2 \exp(-\varepsilon^2 n^{1-2\beta}/4)+n C \exp(-s n^{\beta}/2),
\end{align}
We then deal with $\hat{I}_{l,2}$ and $\hat{I}_{l,3}$. It is easy to have that
\begin{align*}
\abs{\hat{I}_{l,2}}&\leq  \sum_k \hat{p}_k\abs{U_{n_k, l}- \Delta_{l|k}}\leq \max_{k}\abs{U_{n_k, l}- \Delta_{l|k}},\\
\abs{\hat{I}_{l,3}} &\leq \sum_{k=1}^{K_n} \abs{\hat{p}_k-p_k} \Delta_{l|k} \leq K_n \max_k \{\abs{\hat{p}_k-p_k}\Delta_{k,l}\}\leq C K_n \max_k \abs{\hat{p}_k-p_k}.
\end{align*}
This with (\ref{l-l}) results in 
\begin{align}
P(\abs{\hat{I}_{l,2}} \geq \varepsilon)&\leq P( \max_{k}\abs{U_{n_k, l}- \Delta_{l|k}}\geq \varepsilon)\nonumber\\
&\leq K_n P(\abs{U_{n_k, l}- \Delta_{l|k}}\geq \varepsilon)\nonumber\\
&\leq 2 K_n \exp(-\varepsilon^2 n_k^{1-2\beta}/4)+n_k K_n C \exp(-s n_k^{\beta}/2). \label{I2}
\end{align}
%
and 
\begin{align}
P(\abs{\hat{I}_{l,3}} \geq \varepsilon)&\leq P(C K_n \max_{k}\abs{\hat{p}_k-p_k}\geq \varepsilon)\nonumber
=P( \max_{k}\abs{\hat{p}_k-p_k}\geq  \varepsilon/(CK_n))\nonumber\\
& \leq  2K_n \exp\{-\dfrac{n\varepsilon^2/(CK_n)^2}{2(p_k+\varepsilon/(3CK_n))}\} \label{I30} \\
&\leq  2K_n \exp\{-C \dfrac{n}{K_n}\varepsilon^2\}. \label{I3}
\end{align}
The inequality (\ref{I30}) is from Lemma \ref{kk} and the last inequality is derived by the condition \textbf{C1}.

Combining (\ref{I1}), (\ref{I2}) and (\ref{I3}), we have
\begin{align*}
&P(\abs{\hat{\mbox{gCov}}(\bi X_{(l)}, Y)-\mbox{gCov}(\bi X_{(l)}, Y)}\geq \varepsilon) \\
& \leq P(\abs{\hat{I}_{l,1}} \geq \varepsilon/3)+P(\abs{\hat{I}_{l,2}} \geq \varepsilon/3)+P(\abs{\hat{I}_{l,3}} \geq \varepsilon/3)\\
&\leq 2 \exp(-\varepsilon^2 n^{1-2\beta}/36)+n C \exp(-s n^{\beta}/2)\\
&+2K_n  \exp(-\varepsilon^2 n^{1-2\beta}/36)+n CK_n \exp(-s n^{\beta}/2)+ 2K_n \exp\{-C \dfrac{n}{K_n}\varepsilon^2\} .\\
&\leq 2K_n  \exp(-\varepsilon^2 n^{1-2\beta}/36)+n C K_n \exp(-s n^{\beta}/2)+ 2K_n \exp\{-C \dfrac{n}{K_n}\varepsilon^2\} .
\end{align*}

\noindent{\textbf{Proof of Theorem \ref{ucb_gCov}.}}

The Gini distance correlation, $\hat{\rho_l}$,  has the same convergence rate as $\hat{\mbox{gCov}}(\bi X_{(l)}, Y), l=1,2,...,r$.
Let $\varepsilon=cn^{-\kappa}$, for $0<\kappa+\beta< 1/2$, under the condition that $K_n=O(n^\tau)$, we have
\begin{align*}
&P\big(\max_{1 \leq l \leq r}\abs{\hat{\rho_l}-\rho_l} \geq c n^{-\kappa}\big) \leq r \max_{1 \leq l \leq r}P\big(\abs{\hat{\rho_l}-\rho_l} \geq c n^{-\kappa}\big)\\
&\leq O\left\{r\left[K_n  \exp(-b_1 n^{1-2(\kappa+\beta)})+n K_n  \exp(-b_2 n^{\beta})+ 2K_n \exp\{-C \dfrac{n}{K_n} n^{-2 \kappa}\}\right ]\right\}\\
&\leq  O\left\{r\left[  \exp(-b_1 n^{1-2(\kappa+\beta)}+\tau \log n)+  \exp(-b_2 n^{\beta}+(1+\tau)\log n)+ \exp(-b_3 n^{1-2 \kappa-\tau}+\tau\log n)\right ]\right\},
\end{align*}
which has shown the inequality (\ref{uni_con}). The last inequality is by the condition \textbf{C1} that $K_n=O(n^{\tau})$.

If ${\cal A} \nsubseteq \hat{\cal A}$, then there must exist some $l \in {\cal A}$ such that $\hat{\rho}_l<cn^{-\kappa}$. By condition \textbf{C2}, $\abs{\hat{\rho}_l-\rho_l}\geq c n^{-\kappa}$ for some $l \in {\cal A}$. This implies that 
$\{{\cal A} \nsubseteq   \hat{\cal A}\} \subseteq \{|\hat{\rho}_l-\rho_l|>cn^{-\kappa}, \text{for some $l \in {\cal A}$}\}$ and hence ${\cal B}_n=\{\max_{l \in {\cal A}}|\hat{\rho}_l-\rho_l| \leq cn^{-\kappa}\} \subseteq\{ {\cal A} \subseteq \hat{\cal A}\}$.
\begin{align*}
P( {\cal A} \subseteq \hat{\cal A}) &\geq P({\cal B}_n)= 1-P({\cal B}^c_n)
=1-P(\min_{l \in {\cal A}}|\hat{\rho}_l-\rho_l| \geq cn^{-\kappa})\\
&=1-d_nP(|\hat{\rho}_l-\rho_l| \geq cn^{-\kappa})\\
&\geq 1-O \bigg(d_n\left[  \exp(-b_1 n^{1-2(\kappa+\beta)}+\tau \log n)+  \exp(-b_2 n^{\beta}+(1+\tau)\log n)\right.\\
&\left.+ \exp(-b_3 n^{1-2 \kappa-\tau}+\tau\log n)\right ]\bigg),
\end{align*}
where $d_n$ is the cardinality of ${\cal A}$. We have completed the proof of (\ref{screening}). 

\noindent\textbf{Proof of Theorem \ref{rc}.}
We will prove Theorem \ref{rc} by showing that 
\begin{align*}\label{sup}
P\bigg\{{\lim \sup}_{n \to \infty}\big(\min_{l \in \mathcal{A}}\hat{\rho}_l-\max_{l \in \mathcal{I}}\hat{\rho}_l\big)<c_3/2 \bigg\}=0.
\end{align*}
By condition \textbf{C4} and Lemma \ref{lemma4}, we have
\begin{align*}
P\bigg\{\big(\min_{l \in \mathcal{A}}\hat{\rho}_l-\max_{l \in \mathcal{I}}\hat{\rho}_l\big ) < c_3/2\bigg\}& \leq P\bigg\{\big(\min_{l \in \mathcal{A}}\hat{\rho}_l-\max_{l \in \mathcal{I}}\hat{\rho}_l\big )- \big(\min_{l \in \mathcal{A}}\rho_l-\max_{l \in \mathcal{I}}\rho_l\big )< -c_3/2\bigg\}\\
& \leq P\bigg\{\abs{\big(\min_{l \in \mathcal{A}}\hat{\rho}_l-\max_{l \in \mathcal{I}}\hat{\rho}_l\big )- \big(\min_{l \in \mathcal{A}}\rho_l-\max_{l \in \mathcal{I}}\rho_l\big )}>c_3/2\bigg\}\\
& \leq  P\bigg\{2 \max_{1 \leq l \leq r}\abs{\hat{\rho}_l-\rho_l}>c_3/2\bigg\}\\
& \leq 2 r K_n  \exp(-b_4 n^{1-2\beta})+n  r K_n \exp(-b_5 n^{\beta}/2)+r K_n \exp(-b_6 \dfrac{n}{K_n})
\end{align*}
for some positive constants $b_4, b_5$, and $b_6$.
By Borel Contelli Lemma, it suffices to show that 
\begin{align*}
\sum_{n=1}^{\infty}\bigg\{2r K_n  \exp(-b_4 n^{1-2\beta})+nr K_n \exp(-b_5 n^{\beta}/2)+r K_n \exp(-b_6 \dfrac{n}{K_n}) \bigg\}< \infty.
\end{align*}
In fact, by condition \textbf{C4}, $\dfrac{K_n \log p }{n^\alpha}=o(1)$ and $\dfrac{K_n \log n}{n^\alpha}=o(1)$, $\alpha=\min\{1-2\beta+\tau, \beta+\tau, 1\}$ we have 
\begin{align*}
&r \leq p \leq \exp(\dfrac{C n^{\alpha}}{2K_n}) \leq \min \{\exp(b_4 n^{1-2\beta}/2), \exp(b_5 n^{\beta}/2), \exp(b_6 n^{1-\tau}/2)\},\\
& \log K_n \leq  \log n, \quad 3 \log n \leq \min\{n^{1-2\beta},\dfrac{n}{K_n}\},\quad 4 \log n \leq n^{\beta}
\end{align*}
for large $n$.
Therefore, there exists $n_0$ such that 
\begin{align*}
&\sum_{n=n_0}^\infty \bigg\{2r K_n  \exp(-b_4 n^{1-2\beta})+nr K_n \exp(-b_5 n^{\beta}/2)+r K_n \exp(-b_6 \dfrac{n}{K_n})  \bigg\}\\
& \leq \sum_{n=n_0}^\infty \bigg\{  \exp\bigg( \log K_n-b_4 n^{1-2\beta}/2\bigg)+ \exp\bigg(\log (nK_n)- b_5 n^{\beta}/2\bigg)+ \exp\bigg( \log K_n-b_6 n^{1-\tau}/2\bigg) \bigg\}\\
& \leq \sum_{n=n_0}^\infty \bigg\{ 2 \exp\bigg( \log K_n-3 \log n\bigg)+ \exp\bigg(\log (nK_n)- 4 \log n\bigg) \bigg\}\\
&\leq 3 \sum_{n=n_0}^ {\infty} n^{-2} < \infty.
\end{align*}
This completes the proof.

\vspace{0.5cm}
\noindent{\bf Acknowledgement} 

\vspace{0.2cm}
\noindent Thanks Wei Zhong for sharing R codes of the MV method and for sharing the Lung Cancer data with us.

\end{document}